\documentclass[12pt]{article}
\pdfoutput=1
\usepackage{pstricks,slashed}
\usepackage{color,verbatim}
\usepackage{amssymb,amsmath,bm,bbm,accents}
\usepackage{epsf,epsfig}
\usepackage{afterpage}
\usepackage{longtable}
\usepackage{cite}
\usepackage{latexsym,mathrsfs,dsfont}
\usepackage{graphics,graphicx}
\usepackage{url}
\usepackage{paralist}
\usepackage{bbold}

\usepackage[T1]{fontenc}
\usepackage[utf8]{inputenc}
\usepackage[english]{babel}
\usepackage{braket}
\usepackage{amsthm}
\usepackage{amsfonts}
\usepackage{float}

\newcommand{\be}{\begin{equation}}
\newcommand{\ee}{\end{equation}}
\newcommand{\bi}{\begin{itemize}}
\newcommand{\ei}{\end{itemize}}
\newcommand{\ba}{\begin{array}}
\newcommand{\ea}{\end{array}}
\newcommand{\bea}{\begin{eqnarray}}
\newcommand{\eea}{\end{eqnarray}}
\newcommand{\dd}{\displaystyle}
\newcommand{\nn}{\nonumber}

\newcommand{\azione}{\mathcal{S}}
\newcommand{\lagrangiana}{\mathcal{L}}

\begin{document}

\begin{flushright} {BARI-TH/21-730}\end{flushright}

\medskip

\begin{center}
{\Large  Chaotic dynamics of a suspended string \\ \vskip 0.3cm
 in a gravitational background with  magnetic field}
\\[1.0 cm]
{ {P.~Colangelo$^a$,   F.~Giannuzzi$^a$ and N.~Losacco$^{a,b}$}
 \\[0.5 cm]}
{\small 
$^a$Istituto Nazionale di Fisica Nucleare, Sezione di Bari,  Via Orabona 4, I-70126 Bari, Italy \\[0.1 cm]
$^b$Dipartimento Interateneo di Fisica ``M. Merlin'', Universit\`a  e Politecnico di Bari, \\ via Orabona 4, 70126 Bari, Italy
}
\end{center}

\vskip 0.8cm

\begin{abstract}
\noindent
We study the effects of a  magnetic field on the chaotic dynamics of a string with endpoints on the boundary of an asymptotically AdS$_5$ space with black hole. We study Poincar\'e sections and compute the Lyapunov exponents for the string  perturbed from the static configuration, for two different orientations, with position of the endpoints on the boundary orthogonal and parallel to the magnetic field. We find that the magnetic field stabilizes the string dynamics, with the largest Lyapunov exponent remaining below the Maldacena-Shenker-Stanford bound.  
\end{abstract}

\thispagestyle{empty}


\section{Introduction}
The aim of this paper is to study the effect of a constant and uniform magnetic field on the chaotic behavior of a suspended string in a gravitational background.
Among others, one purpose is to scrutinize the stabilization role of the magnetic field for different orientations of the string. The study follows the tests, carried out using holographic methods, of the Maldacena-Shenker-Stanford (MSS) bound \cite{Maldacena:2015waa}. This conjectures, under general conditions, that for a  thermal quantum system  at temperature $T$ some out-of-time-ordered correlation functions involving  Hermitian operators have an exponential  time dependence in determined time intervals.  The dependence is  characterized by the exponent  $\lambda$,  and for such an exponent an upper bound  (written in units where  $\hbar=1$ and $k_B=1$) holds:
\begin{equation}
\lambda \leqslant 2 \pi T .
\label{eq:1}
\end{equation}
The correlation functions are related to  thermal expectation values of the squared commutator of two Hermitian operators at a time separation $t$, which quantify the effect of one operator on  measurements of the other one at a later time. 

The MSS bound  has been inspired by the observation that in nature the black holes (BH)
are the fastest ``scramblers'': the time needed for a system near a BH horizon to loose  information depends  logarithmically on the number of the system degrees of freedom  \cite{Sekino_2008, susskind2011addendum}. Connections between chaotic quantum systems and gravity  have been investigated in  \cite{Shenker_2014,Shenker:2014cwa,Kitaev,Polchinski:2015cea}. In a holographic framework,  a relation has  been worked out  between the size of the operators of the quantum theory on the boundary, which are involved in the temporal evolution of the perturbation, and the momentum of a particle falling in the bulk \cite{Susskind:2018tei,Brown:2018kvn}.

Holographic methods have been used to challenge  the MSS bound \eqref{eq:1}. In such studies the quantum system is conjectured to be a $4d$ boundary theory  dual to an AdS$_5$ gravity theory with a black hole \cite{Maldacena_1998, Witten:1998qj, Gubser_1998}. 
Several investigations  are described in \cite{deBoer:2017xdk,Dalui:2018qqv,Dalui:2020qpt,Ageev:2021xgy}.
Some studies  concern strings hanging in the bulk  with endpoints on the boundary, 
which are the  holographic dual of a static  quark-antiquark pair  \cite{Avramis:2006nv,Arias_2010,Nunez:2009da,Bellantuono:2017msk}.
In such systems  $\lambda$ is  the Lyapunov exponent characterizing the chaotic behavior of the  fluctuations around the static string  configuration \cite{Hashimoto:2018fkb, Ishii_2017,Akutagawa_2019}. The studies include the case of quantum systems characterized by a global $U(1)$ symmetry, and show that the chemical potential stabilizes the chaotic dynamics \cite{Colangelo:2020tpr}.

It is interesting to consider the role of an external magnetic field on the chaotic behaviour of the string. The magnetic field is relevant in different contexts, 
including  heavy-ion collisions or condensed matter problems such as the Quantum Hall Effect and superconductivity at high temperatures. A general gravity dual  for  such systems should include a magnetic field \cite{Critelli:2016cvq,Ballon-Bayona:2020xtf,Arefeva:2020vae,Arefeva:2020bjk,Arefeva:2021mag}.
The backreaction of an external magnetic field modifies the geometry of the $5d$ spacetime, the metric of which is determined by the Einstein equations. 
As a result, an anisotropy  is  introduced in the spatial directions. 
Moreover, in a finite temperature system the relation between the position of the black-hole horizon, the source of chaos in the $5d$ geometry, and temperature,  involved in the MSS relation in the boundary theory, is modified by the magnetic field.

In this paper we aim at  studying how  the background magnetic field affects chaos for the hanging string, how this  depends on the string orientation, and if the MSS bound is satisfied.

\section{Geometry with a magnetic field}\label{geometry}
\begin{figure}[t!]
\begin{center}
\includegraphics[width=0.5 \textwidth]{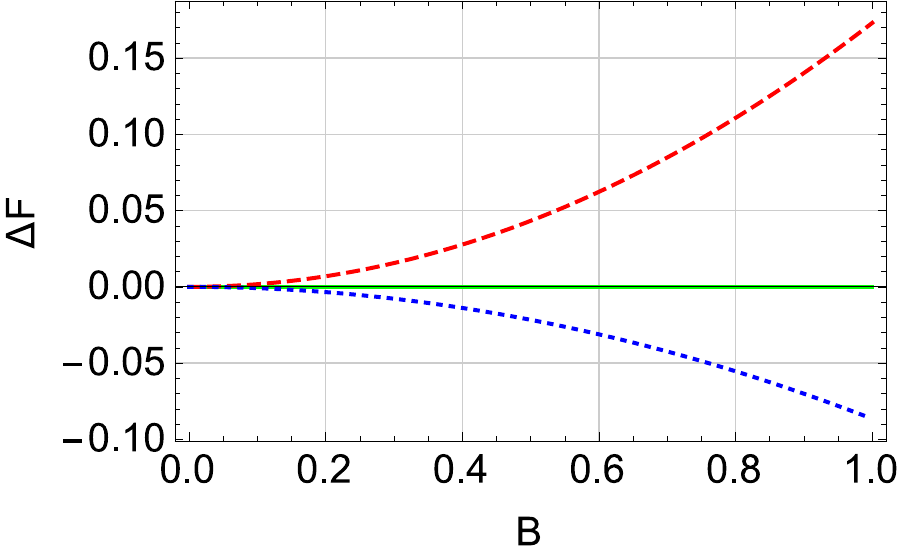}
\caption{\small For $F$  one of the metric functions $f(r=1.1)$  (green line), $q(r=1.1)$ (red dashed line) and $h(r=1.1)$  (blue dotted line),  $\Delta F$ is the difference between each function  in Eqs.~\eqref{eq:gtt}-\eqref{eq:g22} and the corresponding one computed in Ref.\cite{DHoker:2009mmn,DHoker:2009ixq,Braga:2021fey}, for $r_h=1$ and  varying the magnetic field $B$.}
\label{fig:huang-braga}
\end{center}
\end{figure}

\begin{figure}[t!]
\begin{center}
\includegraphics[width=0.5 \textwidth]{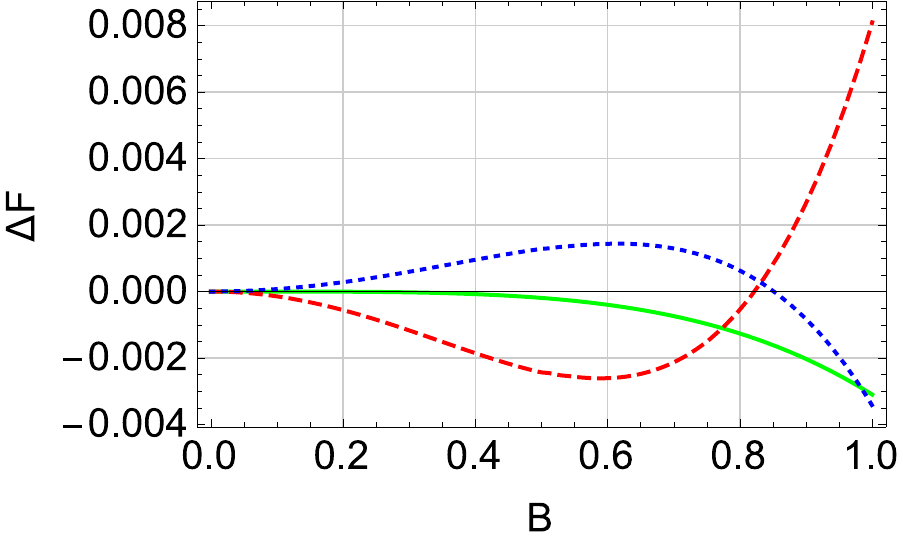}
\caption{\small For $F$ one of the metric functions $f(r=1.1)$  (green line), $q(r=1.1)$ (red dashed line) and $h(r=1.1)$  (blue dotted line),  $\Delta F$ is the difference between each function  in \eqref{eq:gtt}-\eqref{eq:g22} and the one numerically computed from the Einstein equations, for $r_h=1$ and varying the magnetic field $B$.}
\label{fig:huang-numerical}
\end{center}
\end{figure}

In the gauge/gravity duality, a $4d$ boundary gauge theory at finite temperature is dual to a gravity theory in  AdS$_5$ with a black hole. 
A magnetic field is introduced in the holographic framework by a $U(1)$ gauge field $F_{MN}$ which modifies the $5d$ geometry.
The metric is determined  solving the Einstein equations:
\begin{equation}
    R_{MN}-\frac{1}{2} g_{MN} (R+12)-T_{MN}=0 
    \label{eq:einstein}
\end{equation}
with the $5d$ stress-energy tensor
\begin{equation}
    T_{MN}= 2 \, (g^{AB} F_{MA} F_{NB} -\frac{1}{4} g_{MN} F^2) \,.
    \label{eq:stress}
\end{equation}
For a constant magnetic field $B$ in the $x_3$ direction $F$ is given by $F=B \, dx^1\wedge dx^2$, hence the only nonvanishing components are $F_{12}=-F_{21}=B$.
The Einstein equations have been solved perturbatively in the low-$B$ and high temperature limits in Refs.\cite{DHoker:2009mmn,DHoker:2009ixq,Li:2016gfn}. The result for the line element, having the general expression
\begin{equation}\label{eq:metricHuang}
    ds^2=g_{tt} dt^2 + g_{11}(dx^1)^2+ g_{22} (dx^2)^2 +g_{33} (dx^3)^2+g_{rr} dr^2 
\end{equation}
with $r>r_h$,  reads:  
\be \label{eq:metricHuang1}
g_{tt}=-r^2 f(r) , \quad g_{11}=g_{22}=r^2 h(r),  \quad g_{33}=r^2 q(r),  \quad g_{rr}=\frac{1}{r^2 f(r)}. 
\ee
The metric functions are \cite{Li:2016gfn}:
\begin{eqnarray}
    f(r)&=&1-\frac{2B^2}{3 r^4}  \log r+\frac{f_4}{r^4} \label{eq:gtt}\\
    q(r)&=&1-\frac{2 B^2}{3 r^4}  \log r \label{eq:g33} \\
    h(r)&=&1+\frac{B^2}{3 r^4}  \log r\,. \label{eq:g22}
\end{eqnarray}
The magnetic field breaks rotational invariance, hence  $g_{22}\neq g_{33}$.
The geometry has a horizon, the position of which $r_h$  is found  requiring $f(r_h)=0$. This gives $\dd f_4 = - r_h^4 + \frac{2}{3}\, B^2 \log(r_h)$
and the blackening function $f(r)$:
\be
f(r)=1-\frac{r_h^4}{r^4}-\frac{2 B^2}{3 r^4}  \log {\frac{r}{r_h}} . \label{eq:gtt1}
 \ee
The  Hawking temperature $T$ depends on the magnetic field: 
\begin{equation}\label{eq:TvsrhB}
    T = \frac{r_h}{\pi} \left(1 - \frac{B^2}{6 r_h^4}\right) \,.
\end{equation}
The metric given in terms of the functions \eqref{eq:gtt}-\eqref{eq:g22}  is  obtained  for large bulk coordinate $r$ and low $B$,  and it is important to reckon the minimum value of $r$ and the largest value of $B$ for which it is a good approximation of Eqs. \eqref{eq:einstein},\eqref{eq:stress} and \eqref{eq:metricHuang1}.
In Fig.~\ref{fig:huang-braga} the differences between the metric functions $f(r)$, $q(r)$ and $h(r)$  in \eqref{eq:gtt}-\eqref{eq:g22} and the corresponding  ones computed for low $B$ in \cite{DHoker:2009mmn,DHoker:2009ixq} and  used in \cite {Braga:2021fey} are depicted setting  $r=1.1$ and  varying $B$. The comparison shows that the largest deviation  between the two expressions of the metric,  resulting from the different approximations in determining the solution of the Einstein equations, is in the function $q$. For larger values of the coordinate $r$ the deviations are smaller. 
In Fig.~\ref{fig:huang-numerical} the differences between the metric functions in \eqref{eq:gtt}-\eqref{eq:g22} and the ones obtained by a numerical solution of the  Einstein equations are depicted for the same value $r=1.1$,  varying the magnetic field $B$. The numerical solutions are obtained imposing as boundary conditions  for  $r\to \infty$ the asymptotic expansion in \eqref{eq:gtt}-\eqref{eq:g22}, plus additional terms up to $\mathcal{O}(1/r^{8}$). The parameter $f_4$  in the  asymptotic functions is fixed imposing $f(r_h)=0$. 
The deviations between \eqref{eq:gtt}-\eqref{eq:g22} and those in Refs.\cite{DHoker:2009mmn,DHoker:2009ixq,Braga:2021fey} are larger because in such  references  different conditions for $q(r_h)$ and $h(r_h)$  are imposed.
In view of this comparison, in our study  the magnetic field is increased up to  $B=1$. For the radial coordinate $r$, setting the horizon position $r_h=1$ in all our study,  we consider  $r\geqslant 1.1$.

\section{String profile in the gravitational background}\label{profile}
We consider a string  described by the functions $r(t,\ell)$ and $x_i(t,\ell)$, with fixed endpoints on the AdS boundary $r\to\infty$. We consider two configurations,  $i=1$ and $i=3$.
In the former (latter) configuration the string endpoints lie on a line orthogonal (parallel) to the magnetic field. 
$(t,\ell)$ are the worldsheet coordinates, with $\ell$  the proper distance measured along the string.
In the probe approximation we ignore the backreaction to the metric \eqref{eq:metricHuang}-\eqref{eq:g22}.

\begin{figure}[h]
    \begin{center}
    \includegraphics[width=0.5 \textwidth]{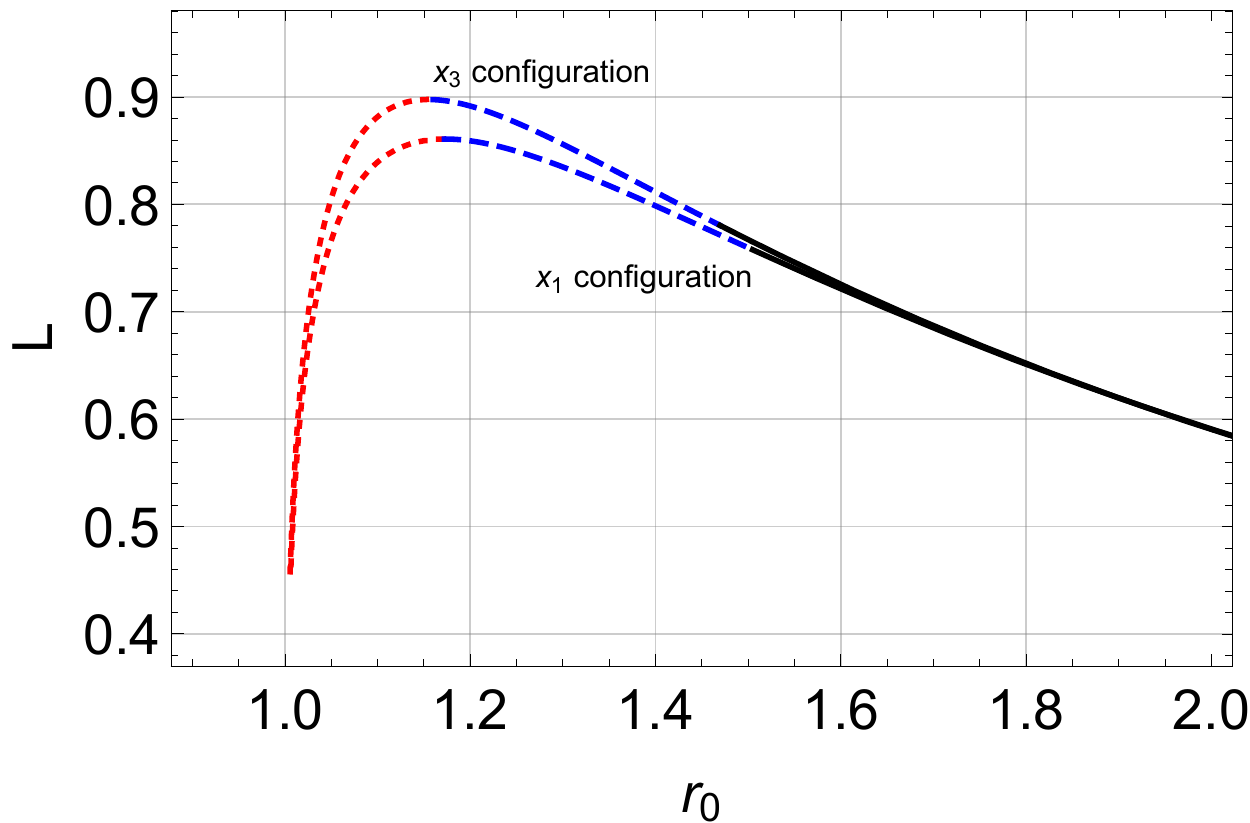}
    \caption{\small Distance between the string endpoints on the boundary vs the position of the tip of the string, obtained using   Eq.~\eqref{eq:Lvsr0} with $r_h=1$ and $B=1$, for the two string configurations.}
    \label{fig:Lvsr0}
    \end{center}
\end{figure}

The string dynamics is governed by the Nambu-Goto (NG)  action:
\begin{equation}
    \azione=-\frac{1}{2\pi \alpha^\prime} \int dt\, d \ell \, \sqrt{-h}\,,
\end{equation}
where $\alpha^\prime$ is the string tension and $h$ is the determinant of the induced metric $h_{ij}=g_{MN} \frac{\partial X^M}{\partial \xi_i} \frac{\partial X^N}{\partial \xi_j}$, with $\xi_{i,j}$ the worldsheet coordinates and $g$ the metric tensor \eqref{eq:metricHuang1}.
In the static case the action reads:
\begin{equation}
    \azione = -\frac{T}{2\pi\alpha^\prime}\int d\ell \, \sqrt{|g_{tt} g_{ii} {\acute x_i}^2 + g_{tt} g_{rr} {\acute r}^2|} \,,
\end{equation}
where $\acute x$ denotes the derivative with respect to $\ell$.
$x_i$ is a cyclic coordinate, so its conjugate momentum
\begin{equation}
\frac{\partial \lagrangiana}{\partial \acute x_i}=-\frac{T}{2\pi\alpha^\prime} \frac{|g_{tt}| g_{ii} {\acute x_i}}{\sqrt{|g_{tt}| g_{ii} {\acute x_i}^2 + |g_{tt}| g_{rr} {\acute r}^2}}
\end{equation}
is a constant of motion.
Denoting with $r(\ell=0) = r_0$ the position of the tip of the string in the bulk, i.e. the point where $\dd \frac{dr}{dx_i}\Big |_{\ell=0}=0$, we have:
\begin{eqnarray}
\frac{\sqrt{|g_{tt}|} g_{ii} {\acute x_i}}{\sqrt{g_{ii} {\acute x_i}^2 + g_{rr} {\acute r}^2}} =\left. \sqrt{|g_{tt}| g_{ii}} \right|_{\ell=0}\,.
\end{eqnarray}
Moreover, from the condition
\begin{eqnarray}
&& d\ell^2=g_{ii}\, dx_i^2+g_{rr}\, dr^2 
\end{eqnarray}
the equations determining the string profile can be obtained:
\begin{eqnarray}\label{eq:eqxstatic}
\acute x=\pm \frac{\sqrt{-g_{tt}(r_0) g_{ii}(r_0)}}{\sqrt{-g_{tt}} g_{ii}} \\
\acute r=\pm \frac{\sqrt{-g_{tt} g_{ii} + g_{tt}(r_0) g_{ii}(r_0) }}{\sqrt{-g_{tt} g_{ii} g_{rr}}}\,.
\label{eq:eqrstatic}
\end{eqnarray}
We set as  boundary conditions that the string  endpoints lie on the AdS$_5$ boundary at $x_i=\pm L/2$. The minimum value $r_0$ of the coordinate $r$ is reached at $x_i=0$ (or $\ell = 0$).  $L$ and $r_0$  are related, since
\begin{equation}\label{eq:Lvsr0}
    L = 2 \int_{r_0}^\infty dr \, \Bigg(\frac{g_{ii}(r)}{g_{rr}(r)} \left( \frac{g_{tt}(r) g_{ii}(r)}{g_{tt}(r_0) g_{ii}(r_0)}-1\right) \Bigg)^{-\frac{1}{2}} \,.
\end{equation}
The function $L(r_0)$ with $B=1$   is plotted in Fig.~\ref{fig:Lvsr0}.  It has a maximum  separating unstable strings (red dotted line in the figure), corresponding to positive energies, from metastable (blue dashed line) and stable strings (black solid line) corresponding to negative energies \cite{Colangelo:2020tpr}.
In the following we  focus on the  unstable string solutions,  varying the magnetic field  in the range $B\leqslant 1$.

\section{Perturbing the static solution}\label{expansion}
To observe the onset of chaos  the static solution of the string near the black-hole horizon must be perturbed by a small time-dependent effect. 
\begin{figure}[b!]
	\centering
	\includegraphics[width=0.5 \textwidth]{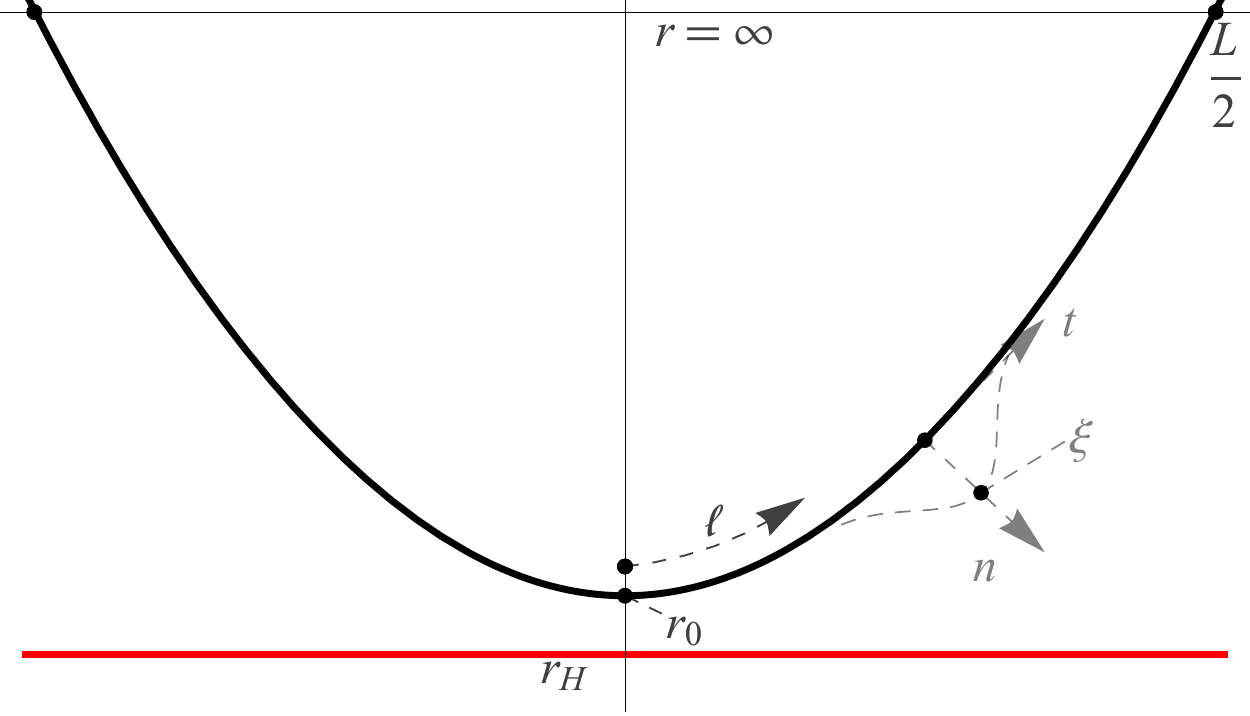}
	\caption[\small Perturbed string]{\baselineskip 12 pt \small Static string profile, and perturbation $\xi(t,\ell)$ along the direction orthogonal to the string in each point having coordinate $\ell$.}
	\label{Fig:1}
\end{figure}
We introduce a perturbation of the string along the orthogonal direction at each point with coordinate $\ell$ in the $r-x$ plane, for both the   $x=x_1$ and  $x=x_3$ configurations \cite{Hashimoto:2018fkb,Colangelo:2020tpr}.  The perturbation is depicted in Fig.~\ref{Fig:1}.
Considering the unit vector $n^M= (0,n^x,0,0,n^r)$ orthogonal to $t^M$, we have:

\bea
g_{rr} ( r ) \left({n}{^{r}}\right)^2 +g_{xx} ( r) \left({n}{^{x}}\right)^2 &=& 1
\label{eq:1M} \\
\acute{r} \left( \ell \right)  g_{rr} \left( r \right)  \, n^r + \acute{x} \left( \ell \right)  g_{xx}\left( r \right)  \, n^x&=&0 \,.
\label{eq:2M}
\eea
For an outward perturbation as  in Fig.~\ref{Fig:1}  the solution for the components $n^{x}$ and $n^{r}$  is

\begin{equation}
{n^x} ( \ell)=\sqrt{\frac{g_{rr}}{g_{xx}}} \; \acute{r}( \ell ) \quad \text{,} \quad {n}{^{r}} ( \ell )=-\sqrt{\frac{g_{xx}}{g_{rr}}} \; \acute{x}( \ell ) \quad .
\label{eq:3}
\end{equation}
\noindent 
The time-dependent perturbation $\xi \left( t, \ell \right)$  modifies $r$ and $x$:
\bea
r \left( t,\ell \right) &=& r_{BG} \left( \ell \right) + \xi \left( t, \ell \right) n^{r} \left( \ell \right), \nn \\
x\left( t,\ell \right) &= & x_{BG} \left( \ell \right) + \xi \left( t,\ell \right) n^{x} \left( \ell \right),
\label{eq:4} \nn
\eea
where $r_{BG} \left( \ell \right)$ and $x_{BG} \left( \ell \right)$  are the static solutions obtained integrating Eqs.~\eqref{eq:eqxstatic} and \eqref{eq:eqrstatic}. 

To describe the dynamics of the small perturbation, we expand the metric function around the static solution $r_{BG}(\ell)$  to the third order in  $\xi \left( t,\ell \right)$.
To this order in $\xi$ the NG action comprises a quadratic and a cubic term.
The  quadratic term  has the form:

\begin{equation}
\begin{aligned}
S^{\left( 2 \right)} = \frac{1}{2 \pi \alpha^\prime}\int \mathrm{d}t \int_{-\infty}^{\infty} \mathrm{d} \ell \left( C_{tt}^{x_i}  \dot{\xi}^2 + C_{\ell \ell}^{x_i}  \acute{\xi}^2  + C_{00}^{x_i} \xi^2\right) \qquad \qquad(i = 1,3) 
\end{aligned}
\label{eq:2nd Order Action}
\end{equation}
\noindent
for the two chosen string orientations. $C_{tt}^{x_i}$, $C_{\ell \ell}^{x_i}$ and $C_{00}^{x_i}$ depend on $\ell$. 
For the geometry in Eq.~\eqref{eq:metricHuang} with metric functions $f(r)$, $h(r)$ and $q(r)$ the coefficients $C_{tt}^{x_i}$, $C_{\ell \ell}^{x_i}$ and $C_{00}^{x_i}$ read:

\begin{equation}
\begin{aligned}
C_{tt}^{x_i} \left( \ell \right)&=\frac{1}{2 r_{BG} \sqrt{f \left( r_{BG} \right) }},\\
C_{\ell \ell}^{x_i} \left( \ell \right)&=  - \frac{1}{4 C_{tt}^{x_i} \left( \ell \right) } , \\
C_{00}^{x_1} \left( \ell \right) &= \frac{1}{8
r_{BG}^3 f(r_{BG})^{3/2} h(r_{BG})^2 }\Big(\bigl(r_0^4 f(r_0) h(r_0) \bigl(2 r_{BG}^2 h(r_{BG}) f^\prime(r_{BG})^2 \\
&+2 f(r_{BG})^2
\bigl(4 h(r_{BG})+r_{BG} h^\prime(r_{BG})\bigr)+r_{BG} f(r_{BG}) \bigl(r_{BG} f^\prime(r_{BG}) h^\prime(r_{BG})\\
&+ 2
 h(r_{BG}) \bigl(f^\prime (r_{BG})-r_{BG} f^{\prime \prime}(r_{BG})\bigr)\bigr)\bigr)-r_{BG}^4 f(r_{BG})^2 \bigl(2 r_{BG} h(r_{BG})
f^\prime(r_{BG})\\
& \bigl(2 h(r_{BG})+r_{BG} h^\prime(r_{BG})\bigr)+f(r_{BG}) \bigl(8 h(r_{BG})^2-r_{BG}^2
h^\prime(r_{BG})^2\\
&+2 r_{BG} h(r_{BG}) \bigl(4 h^\prime(r_{BG})+r_{BG} h^{\prime \prime}(r_{BG})\bigr)\bigr)\bigr)\Bigr) .
\end{aligned}
\label{eq:6}
\end{equation}
\noindent
$C_{00}^{x_3}$ has the same expression of $C_{00}^{x_1}$,  with the metric function $h(r)$ replaced by $q(r)$.
The coefficients depend on $\ell$ through $r_{BG} \left( \ell \right)$. The metric functions $f(r)$, $h(r)$ and $q(r)$ are  defined in Eqs.~\eqref{eq:gtt}-\eqref{eq:g22}.

\begin{figure}[h]
\centering
\includegraphics[width=0.45 \textwidth]{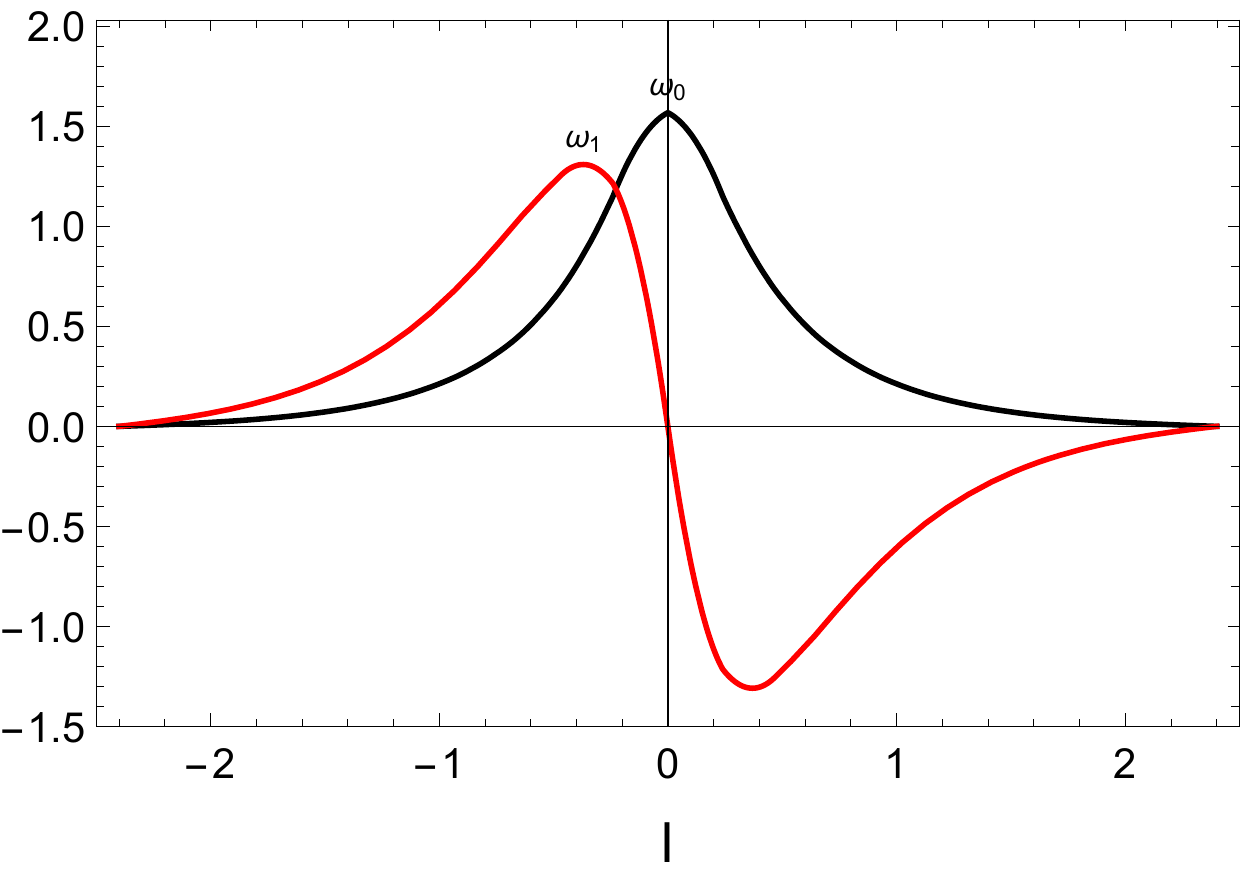}
\includegraphics[width=0.45 \textwidth]{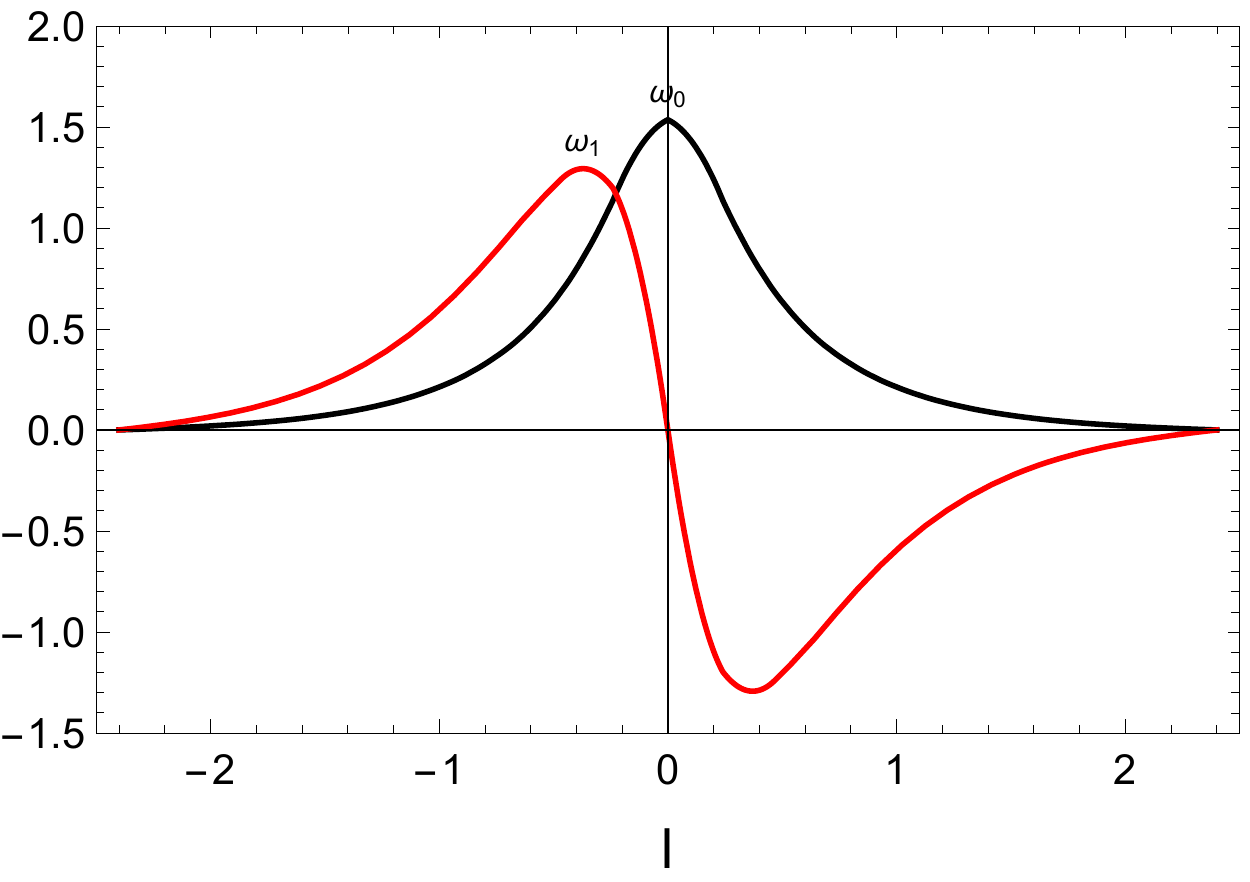}
\caption{\small Eigenfunctions $e_0 \left( \ell \right)$ (black line) and $e_1 \left( \ell \right)$ (red line) of  Eq.~(\ref{eq:9}), for  $r_0=1.1$ and $B=0$ (left),  $B=0.6$ (right panel).}\label{Fig:eigen}
\end{figure}

The  equation of motion from the action \eqref{eq:2nd Order Action} is
\begin{equation}
\qquad \qquad C_{tt}^{x_i} \, \ddot{\xi} + \partial_\ell \left( C_{\ell \ell}^{x_i} \acute{\xi} \right) - C_{00}^{x_i} \, \xi = 0 \qquad \qquad \qquad (i = 1,3).
\label{eq:8}
\end{equation}
Factorizing $\xi \left( t,\ell \right) = \xi \left(\ell \right) e^{i \omega t}$ it  corresponds to the Sturm-Liouville equation 
\be
\partial_\ell \left( C_{\ell \ell}^{x_i} \, \acute{\xi} \right) - C_{00}^{x_i} \, \xi = \omega ^2 C_{tt}^{x_i}  \, \xi \,\, ,
\label{eq:9}
\ee
\noindent
with $W(\ell)=-C_{tt}^{x_i}(\ell)$  the weight function.
We solve Eq.~\eqref{eq:9} for different values of $B$. Since we are interested in unstable configurations we set $r_0 = 1.1$ near the horizon, and  impose the boundary conditions $ \xi \left( \ell \right) \xrightarrow{ \ell \rightarrow \pm \infty } 0$. The  two lowest lying eigenvalues $\omega_0^2$ and $\omega_1^2$, obtained varying $B$ for the two different string configurations, are collected in Table~\ref{Tab:1}.  The corresponding eigenfunctions $\xi \left(\ell \right)=e_0 \left(\ell \right)$ and $\xi \left(\ell \right)=e_1 \left(\ell \right)$, for one configuration of the string, are depicted in Fig.~\ref{Fig:eigen}.
\begin{table}[b!]
\centering
\begin{tabular}{c c c c | c c c c}
\\ \\
\hline \hline
& $x_1$ configuration    			&				&&          & $x_3$ configuration   & 		 \\
\hline \hline 
&$B$       & $\omega_0^2$		& $\omega_1^2$  & &$B$      & $\omega_0^2$		& $\omega_1^2$\\
&0		   & -1.370				& 7.638    		&&0			& -1.370			& 7.638\\
&0.3	   & -1.327				& 7.531			&&0.3		& -1.317			& 7.548\\
&0.6	   & -1.202				& 7.213 		&&0.6		& -1.166			& 7.278\\
&0.9	   & -1.004				& 6.694			&&0.9		& -0.932			& 6.823\\
&1  	   & -0.923				& 6.478 		&&1  		& -0.841			& 6.629\\
\hline
\hline
\end{tabular}
\caption[Values of the first two eigenvalues]{\baselineskip 12 pt \small Eigenvalues $\omega_0^2$ and $\omega_1^2$ of Eq.~\eqref{eq:9} for $r_0 = 1.1$ and varying  $B$, for the two  string configurations.}\label{Tab:1}
\end{table}
\noindent
Negative eigenvalues corrispond to unstable systems. Considering  the results in Table~\ref{Tab:1}, we conclude that the effect of $B$ is to stabilize the system, since $\omega_0^2$  increases with $B$. The effect of the magnetic field $B$ is stronger for the string in the $x_3$ direction, parallel to the magnetic field, hence
  the magnetic field stabilizes the system in the $x_3$ configuration more than in the $x_1$ configuration.

To observe the chaotic behaviour we study the contribution of the  third order terms in $\xi$ in the action. 
Up to a surface term, the  expression  is

\be
S^{\left( 3 \right)} =  \frac{1}{2 \pi \alpha^\prime}\int \mathrm{d}t \int_{-\infty}^{\infty} \mathrm{d}\ell \bigg\{ D_0 \,   \xi^3 
+ D_1^{x_i} \, \xi \acute{\xi}^2 + D_2^{x_i} \, \xi \dot{\xi}^2  \bigg \} \,\,  ,
\label{eq:11}
\ee
with $D_{0,1,2}^{x_i}$ functions of $\ell$.
Expanding the  perturbation in  terms of the first two eigenfunctions  $e_0$ and $e_1$,

\begin{equation}
\xi \left( t,\ell \right) = c_0 \left( t \right) e_0 \left( \ell \right) + c_1 \left( t \right) e_1 \left( \ell \right) ,
\label{eq:10}
\end{equation} 
\noindent 
 the time dependence of the perturbation is encoded in  the coefficients $c_0( t )$ and $c_1 ( t )$. 
With this form of $\xi(t,\ell)$  we have:

\bea
&&S^{\left( 3 \right)}=\nn \\
&& \frac{1}{2 \pi \alpha^\prime}\int \mathrm{d}t \int_{-\infty}^{\infty} \mathrm{d}\ell \Big[ \Big( D_0^{x_i}  \,  e_0^3 + D_1^{x_i} \, e_0 \acute{e}_0^2 \Big) c_0^3 \left( t \right) + \Big( 3 D_0^{x_i}   \, e_0 e_1^2 + D_1^{x_i}  \left( 2 \acute{e}_0 e_1 \acute{e}_1 + e_0 \acute{e}_1^2 \right)\Big) c_0 c_1^2  \nn \\
&&+ D_2^{x_i}  \Big( e_0 e_1^2 c_0 \dot{c}_1^2 +e_0^3 e_1^2 c_0 \dot{c}_0^2+ 2 e_0 e_1^2 \dot{c}_0 c_1 \dot{c}_1    \Big) \Big]. 
\label{eq:12}
\eea
The action for $c_0(t)$ and $c_1(t)$ is obtained by the sum $S^{(2)}+S^{(3)}$, integrating over $\ell$:

\be
S^{(2)}+S^{(3)}=\frac{1}{2 \pi \alpha^\prime}\int \mathrm{d}t \Big[ \sum_{n=0,1}\left(\dot{c}_n^2-\omega_n^2 c_n^2  \right) + K_1^{x_i} c_0^3 
+ K_2^{x_i} c_0 c_1^2  + K_3^{x_i} c_0 \dot{c}_0^2 + K_4^{x_i} c_0 \dot{c}_1^2 + K_5^{x_i} \dot{c}_0 c_1 \dot{c}_1\Big].
\label{eq:13}
\ee
\noindent
\noindent In Eqs. (\ref{eq:11})-(\ref{eq:13})  the index  is $i=1$ or $i=3$.
The  coefficients $K_{1, \dots, 5}^{x_i}$ depend on   $r_0$ and $B$. They are collected in Tab.~\ref{Tab:2} for $r_0 = 1.1$ and different values of  $B$, for the two string configurations.

\begin{table}[b!]
\centering
\begin{tabular}{cccccccc}
\\
\hline
\hline
 $x_1$ configuration \\
\hline
\hline
&{$B$}		& {$K_1$}		& {$K_2$}		&{$K_3$}		&{$K_4$}		&{$K_5$}  \\

&0			& 11.36			& 21.72			& 10.58			& 3.37			& 6.73	      \\
&0.3		& 10.89			& 21.17			& 10.50			& 3.36			& 6.72	      \\
&0.6		& 9.57			& 19.56			& 10.24			& 3.33			& 6.67	      \\
&0.9		& 7.63			& 17.05			& 9.83			& 3.29			& 6.58	      \\
&1			& 6.90			& 16.05			& 9.65			& 3.27			& 6.55	      \\
\hline
\hline
$x_3$ configuration \\
\hline
\hline
&{$B$}		& {$K_1$}		& {$K_2$}		&{$K_3$}		&{$K_4$}		&{$K_5$}  \\

&0			& 11.36			& 21.72			& 10.58			& 3.37			& 6.73	      \\
&0.3		& 10.97			& 21.22			& 10.52			& 3.37			& 6.74	      \\
&0.6		& 9.85			& 19.76			& 10.32			& 3.38			& 6.77	      \\
&0.9		& 8.16			& 17.44			& 9.99			& 3.41			& 6.81	      \\
&1			& 7.50			& 16.50			& 9.85			& 3.41			& 6.83	      \\
\hline
\hline
\end{tabular}
\caption[ $K_{1,\dots,5}$ coefficients of the third order action]{\baselineskip 12 pt \small $K_i$ coefficients in Eq.~(\ref{eq:13}) for the two string configurations, for  $r_0 = 1.1$  and varying the magnetic field $B$.}\label{Tab:2}
\end{table}

The potential described by Eq. \eqref{eq:13} has a trap for the unstable string configurations. We are interested in the motion of $c_0$ and $c_1$ in the trap. In some regions of the  potential the kinetic term is negative.  As shown in \cite{Hashimoto:2018fkb,Colangelo:2020tpr},  it is useful to replace $c_{0,1}\to \tilde c_{0,1}$ in the action, with $c_0=\tilde{c}_0 + \alpha_1 \tilde{c}_0^2 + \alpha_2 \tilde{c}_1^2$ and $c_1 = \tilde{c}_1 + \alpha_3 \tilde{c}_0 \tilde{c}_1$, neglecting  $\mathcal{O} \left( \tilde{c}_i^4 \right)$ terms, setting the constants $\alpha_i$  ensuring the positivity of the kinetic term. We set $\alpha_1=-2$, $\alpha_2 = -0.5$ and $\alpha_3 = -1$. This replacement stretches the potential and stabilizes the time evolution of the system. The dynamics is not affected, and a chaotic behaviour  shows up  also in the transformed system.

\section{Poincar\'e sections and Lyapunov exponents}\label{modes}

\begin{figure*}[t!]
\centering
\makebox[\linewidth][c]{{
	{\includegraphics[width=0.4 \textwidth]{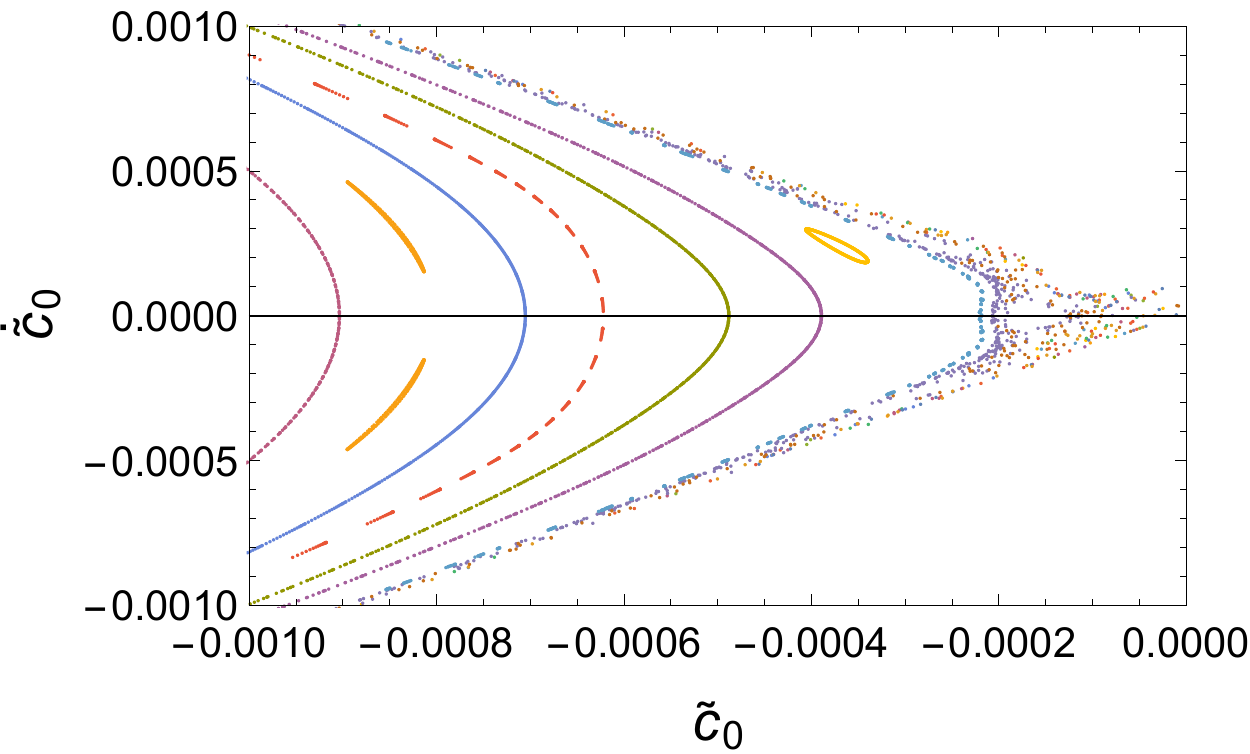} \quad
	 \includegraphics[width=0.4 \textwidth]{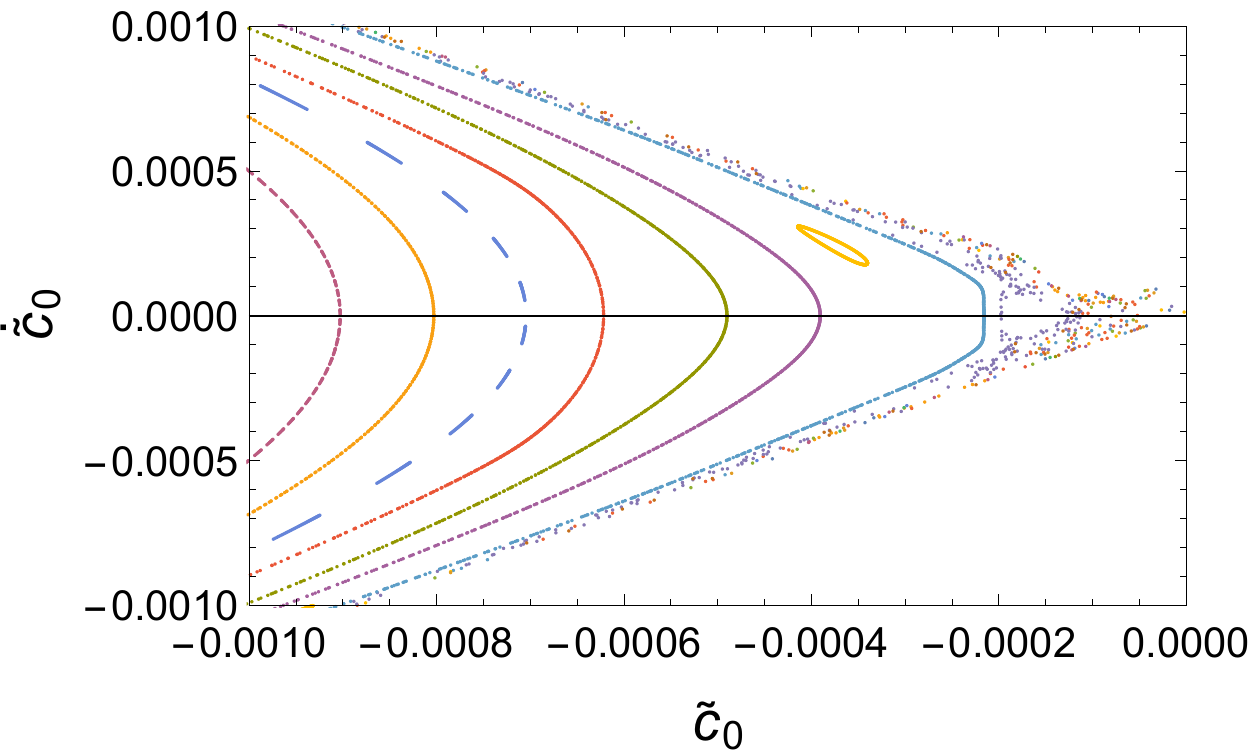}}}} \quad
\makebox[\linewidth][c]{{
	{\includegraphics[width=0.4 \textwidth]{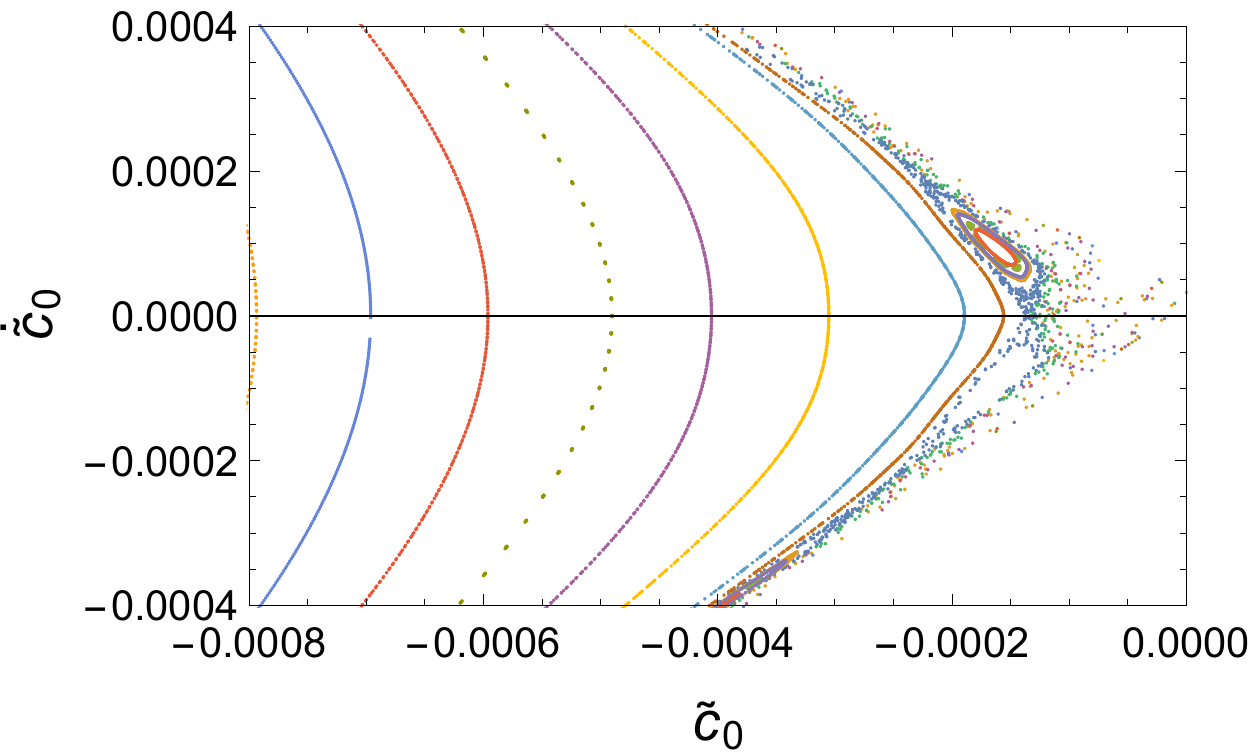} \quad
	 \includegraphics[width=0.4 \textwidth]{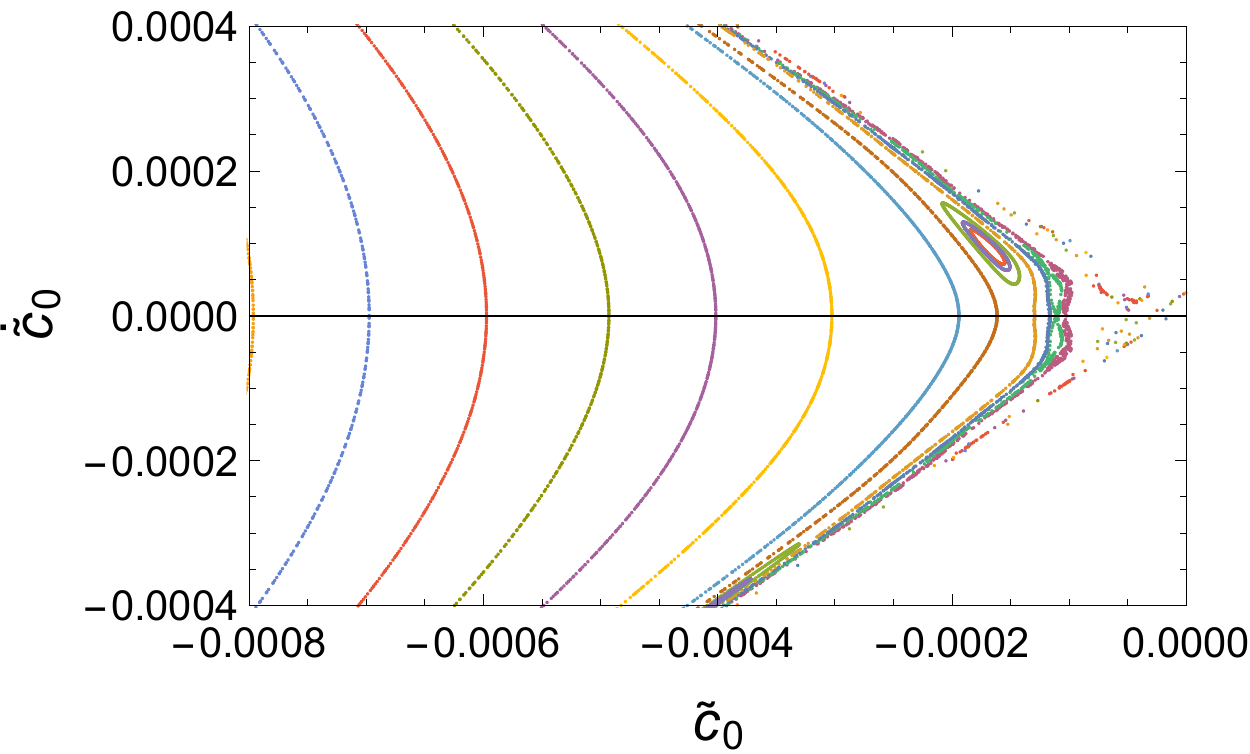}}}}  \quad
\makebox[\linewidth][c]{{
	{\includegraphics[width=0.4 \textwidth]{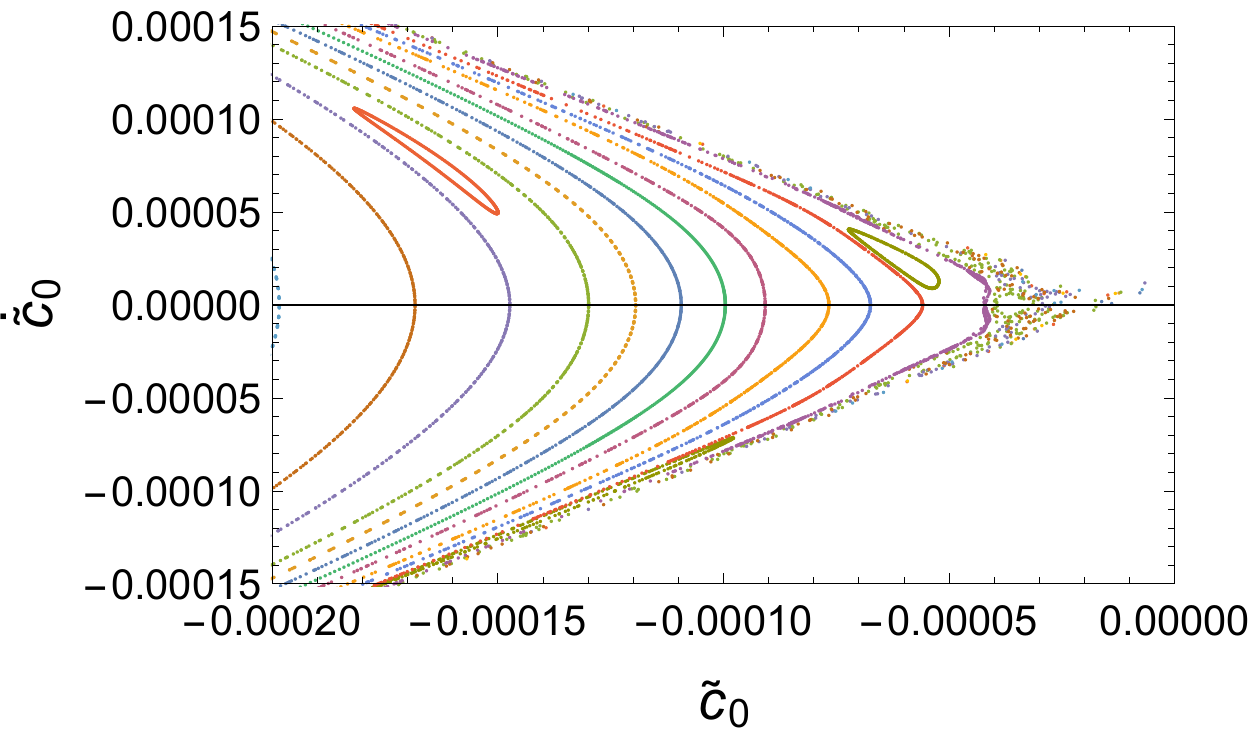}
\quad
	 \includegraphics[width=0.4 \textwidth]{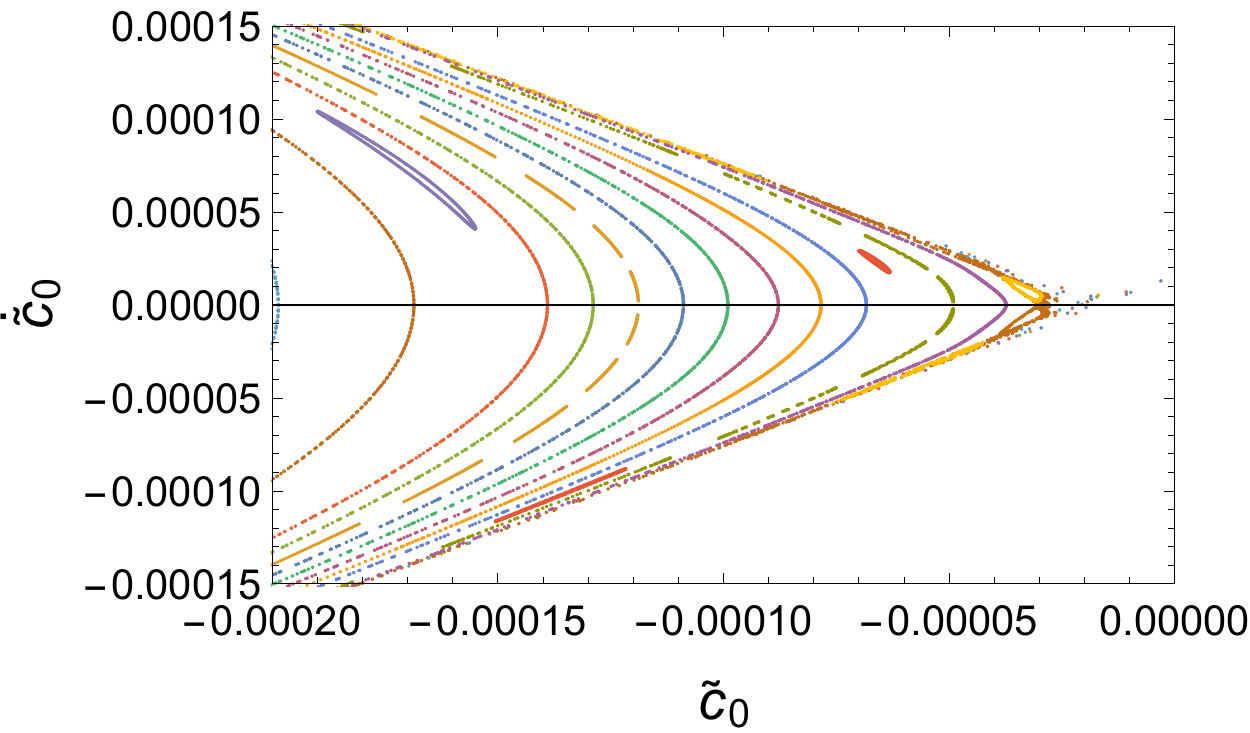}}}}
\quad
\caption[Poincar\'e plots for $r_0 = 1.1$ and different values of $B$]{\small Poincar\'e sections for a perturbed string in the  $x_1$ (left column) and $x_3$ configurations (right column). The initial conditions are changed with fixed energy $E = 10^{-5}$ and $r_0=1.1$. The magnetic field is increased from  $B=0.3$ (top row) to $B=0.6$ (middle row) and $B=1$ (bottom row). The sections  correspond to  $\tilde{c}_1 =0$ and $\dot{\tilde{c}}_1 \geqslant 0$.}
\label{Fig:Poincarè}
\end{figure*}

The onset of chaos is displayed by the  Poincar\'e sections.
We construct the sections defined by $\tilde{c}_1 \left( t \right) = 0$ and $\dot{\tilde{c}}_1 \left( t \right)
\geqslant 0$ for bounded orbits within the trap in the potential. For  $r_0 = 1.1$ and increasing $B$ the sections are depicted  in Fig.~\ref{Fig:Poincarè}. For  $\tilde{c}_0$ near zero the orbits are scattered points which depend on the initial conditions. 
Increasing $B$ the points in the sections form more regular paths, showing that the effect of switching on the magnetic field is to mitigate the chaotic behavior. 

In Fig.~\ref{Fig:Poincarè} we  observe that when the string is along $x_3$ we need to go closer to $\tilde{c}_0 = 0$,  closer to the horizon,  to observe chaos.
This  confirms the observation that the strongest stabilization effect of the magnetic field is in  the  $x_3$ configuration. 

\begin{figure}[t!]
\centering
\includegraphics[width=0.45 \textwidth]{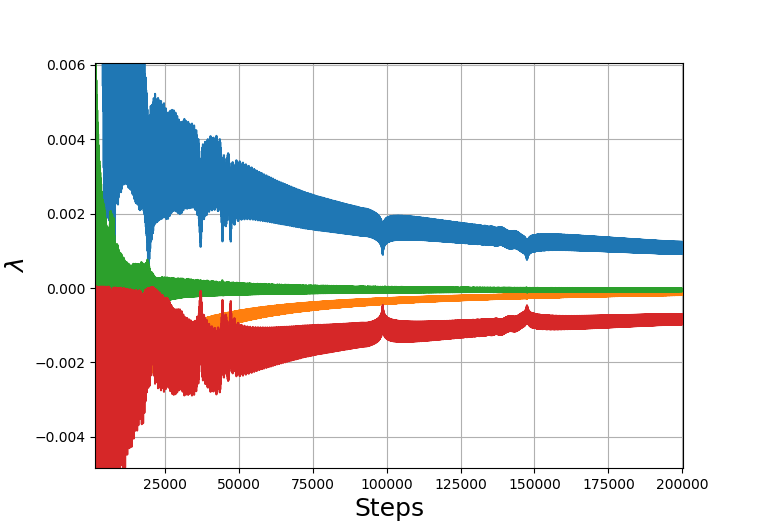}
\includegraphics[width=0.45 \textwidth]{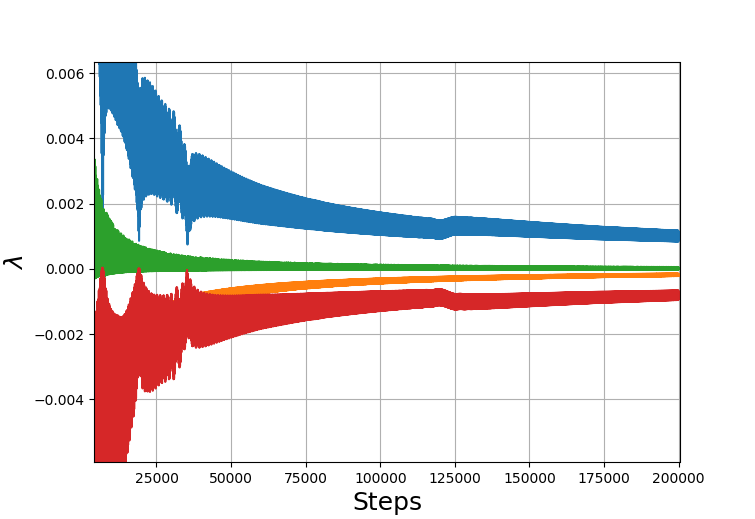}
\caption[Convergency plots of the Lyapunov exponents for a string with $r_0=1.1$]{\baselineskip 12 pt \small Convergency plots of the four Lyapunov exponents 
for  a string along $x_1$, with $r_0=1.1$ and $B=0.3$ (right panel), $B=0.6$ (left panel).  $2\times 10^5$ time steps are shown. For the initial conditions, the energy is set to $E = 10^{-5}$ together with  $\tilde{c}_0 = -0.0002$, $\dot{\tilde{c}}_0 = 0$, $\tilde{c}_1 = 0.0011$.}
\label{Fig:XLCE}
\end{figure}

\begin{figure}[b!]
\centering
	\includegraphics[width=0.55 \textwidth]{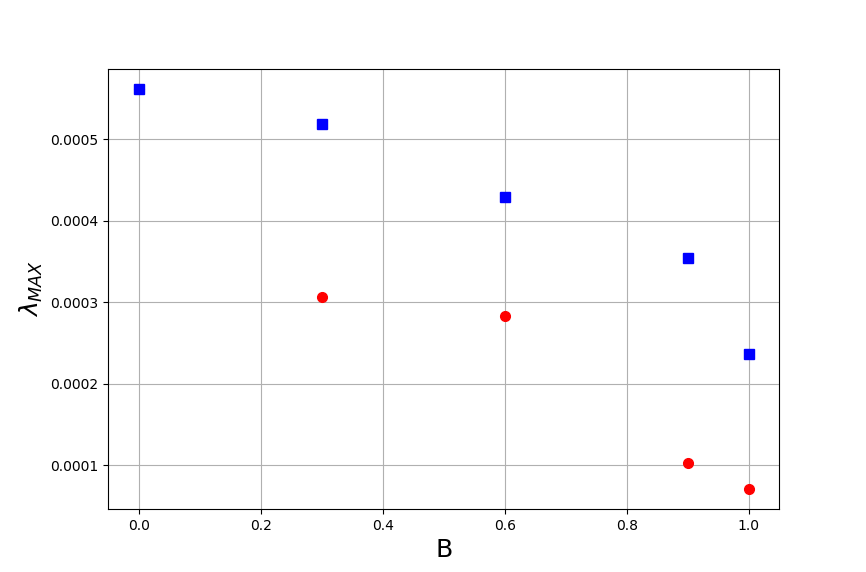} 
	\caption[Largest Lyapunov exponent as a function of $B$ for $x_1$ configuration]{\small Largest Lyapunov exponent  $\lambda_{MAX}$ versus $B$ for $r_0=1.1$. The results for the  $x_1$  (blue points) and $x_3$ string configurations (red points) are shown. }
\label{Fig:XLCEmax}
\end{figure}

For a better understanding of the amount of chaos we  evaluate the Lyapunov exponents.
Such exponents in the four dimensional $c_0$, $c_1$ phase-space can be computed for different values of $B$ using the numerical method described in \cite{sandri1996numerical}. The results are shown in Fig.~\ref{Fig:XLCE}.
\noindent
The convergency plot is a damped oscillating function. The value of the largest Lyapunov exponent can be extrapolated fitting the maximum in each oscillation  and considering  $t \to + \infty$.
The values obtained from the fit decrease as $B$ increases, as shown in Fig.~\ref{Fig:XLCEmax}:  the effect 
of the magnetic field is to soften the dependence on the initial conditions, making the string less chaotic. In the numerical procedure it is checked that the sum of the Lyapunov exponents vanishes at large  $t$.

In Fig.~\ref{Fig:XLCEmax}  the results for the two configurations are compared.
For the same  values of $B$ and $r_0$, so at the same distance from the BH horizon, smaller  Lyapunov exponents are found in the $x_3$ configuration.

\begin{figure}[t!]
    \centering
    \includegraphics[width=0.5 \textwidth]{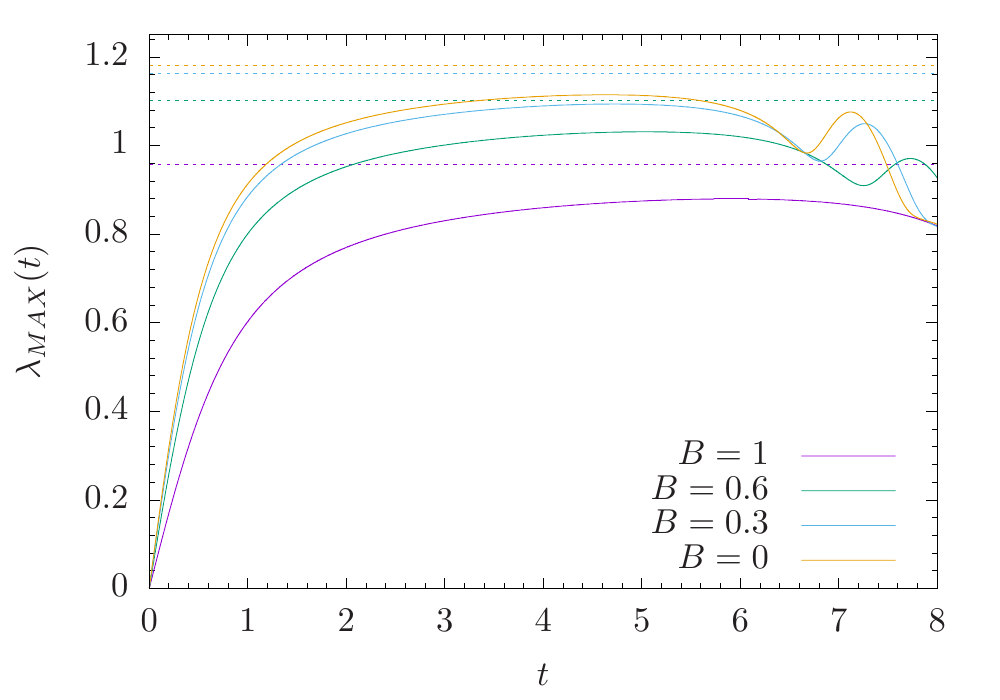}
    \caption{\small Early time convergency plots of the largest Lyapunov exponent, for  the $x_1$ configuration, $r_0=1.1$ and $B$ indicated in the legend. The energy is set to $E = 10^{-5}$, with the initial conditions $\tilde{c}_0 = -0.0002$, $\dot{\tilde{c}}_0 = 0$ and $\tilde{c}_1 = 0.0011$. The dashed lines show the largest Lyapunov exponents at the saddle point  discussed in Sec.~\ref{sec:saddlepoint}.}
    \label{fig:lyapunov-ini}
    \end{figure}

The Poincar\'e plots show that chaos is produced in the proximity of the BH horizon, and that  the string dynamics  is less chaotic if the magnetic field increases. This is confirmed by the largest Lyapunov exponent.
The effect of the BH horizon can  be observed looking at the first steps of the Lyapunov convergency plots shown in Fig.~\ref{fig:lyapunov-ini}.
In the early times, as long as the system is inside a region near $\tilde c_0\sim 0$, the exponents approach a nearly constant value. When the system is  far from the origin, they begin to oscillate and drop to a lower asymptotic value shown in Fig.~\ref{Fig:XLCE}.
This early time behavior is observed when in  general the trajectory crosses a chaotic region of the phase-space. 
If initial conditions are of a trajectory that does not come close to the origin, such a behavior is not observed. 
For the time evolution of the convergency plots  stopped before the plateaux start decreasing, higher values of the largest Lyapunov exponents with respect to the asymptotic  ones in Fig.~\ref{Fig:XLCEmax} would be found.
The snapshot of the time evolution near the BH horizon shown in Fig.~\ref{fig:lyapunov-ini} further shows  that the BH horizon is the source of chaos. The role of  the magnetic field to stabilize the system is also displayed by the  early time behaviour.

\section{Analysis of the saddle point}\label{sec:saddlepoint}
In the previous section we have computed the Lyapunov exponents of an extended bounded orbit in the phase space, and we have obtained positive values for the largest Lyapunov exponents proving that the system is chaotic.
In order to complete the analysis of Lyapunov exponents and challenge the MSS bound, in this section we compute the Lyapunov exponents at the unstable fixed point. 
However, we remark that the sign of the largest Lyapunov exponent at the fixed point only indicates the stability of that  point (it is positive for unstable fixed points), but it carries no information about the dynamics of the whole system and cannot be used to argue if the system is chaotic.

\begin{figure}[t!]
    \begin{center}
    \includegraphics[width=0.6 \textwidth]{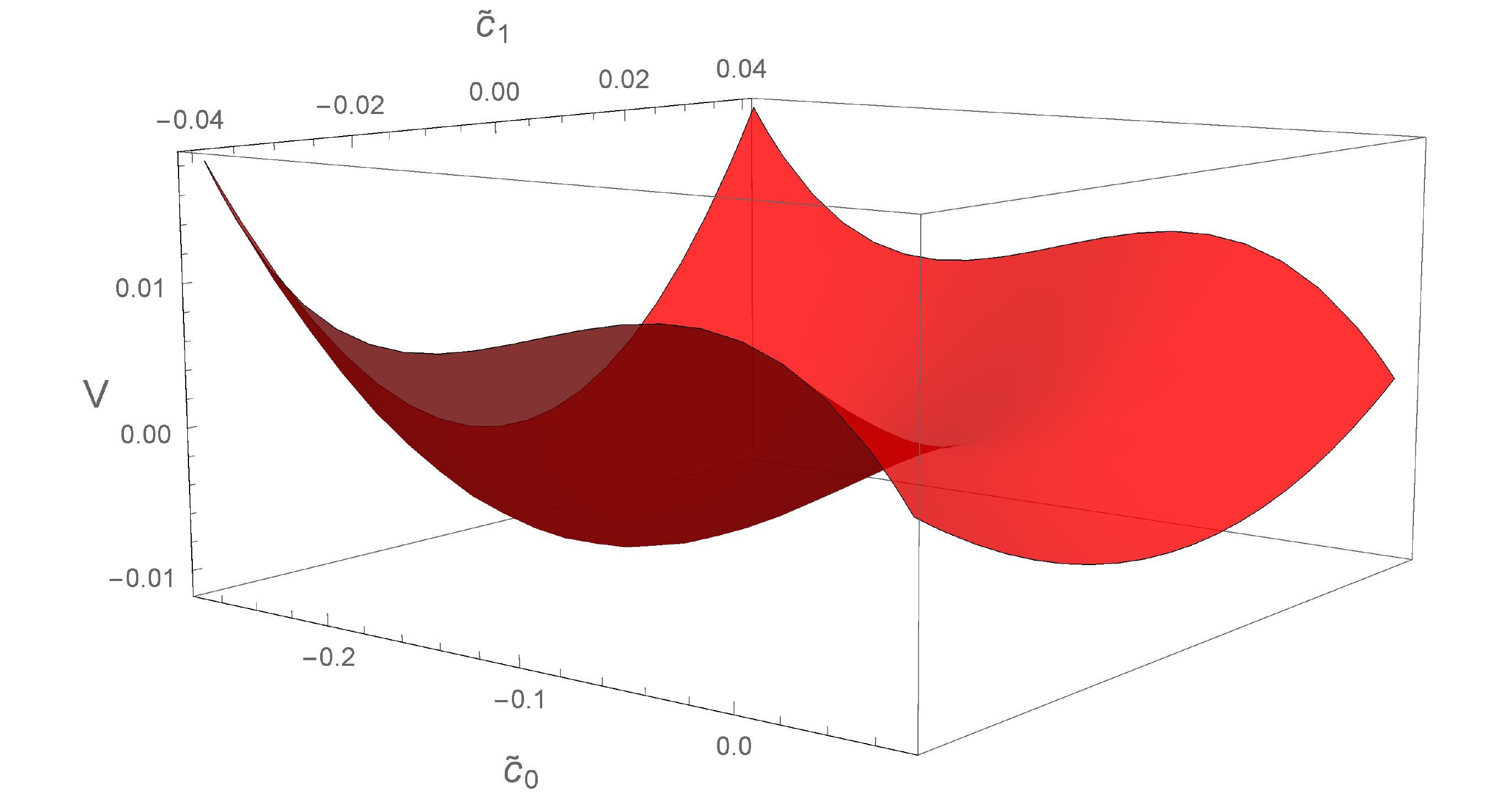}
    \caption{\small Potential obtained from Eq.~\eqref{eq:13} for  $B=1$.}
    \label{fig:pot3d}
    \end{center}
    \end{figure}

The  evolution of the system governed by the action in Eq.~\eqref{eq:13} is dictated by the equation $\dot{\vec x}=\vec F$, with $\vec x=(\tilde{c}_0,\dot{\tilde{c}}_0,\tilde{c}_1,\dot{\tilde{c}}_1)$.
There are two fixed points where $\vec F=0$:  a stable fixed point corresponding to the local minimum of the potential obtained from Eq.~\eqref{eq:13}, and an unstable fixed point corresponding to the saddle point of the potential,  shown in Fig.~\ref{fig:pot3d} for a set of parameters.
    
\begin{figure}[t!]
\begin{center}
\includegraphics[width=0.5 \textwidth]{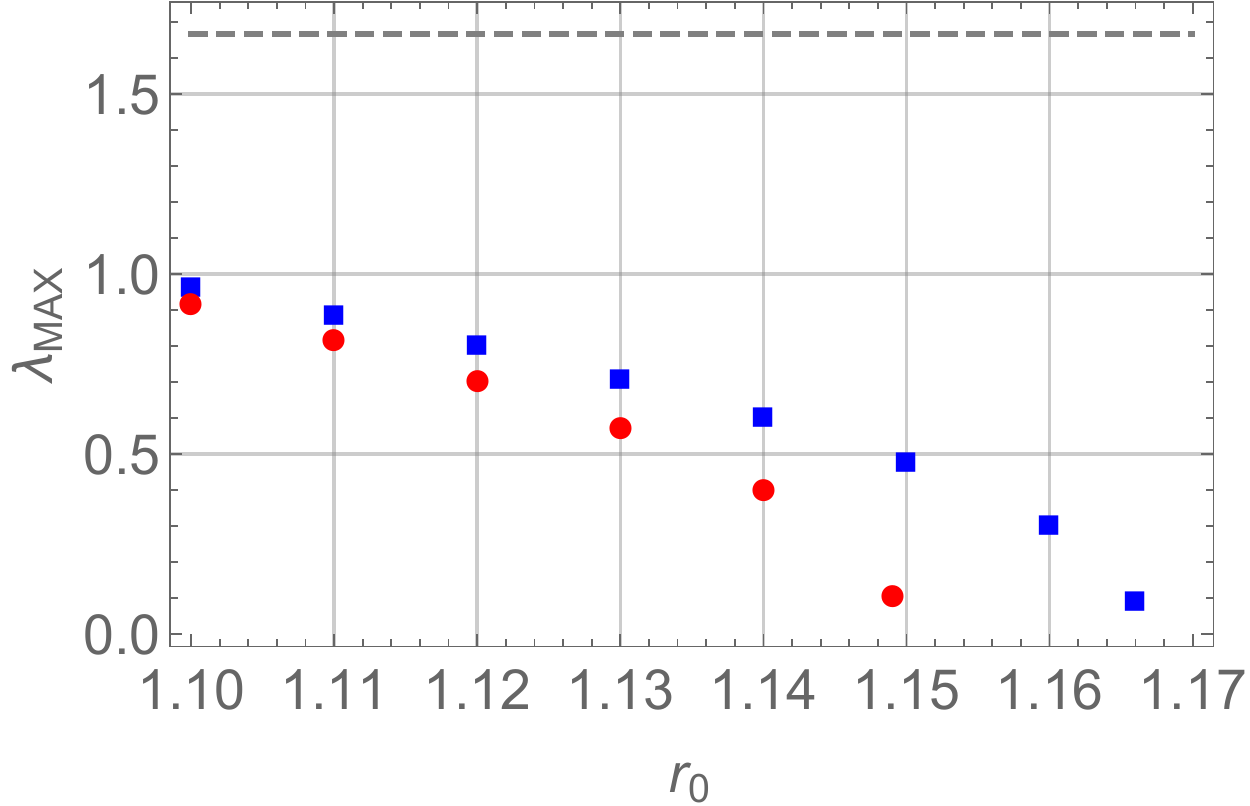} \\
\caption{\small Largest Lyapunov exponent at the unstable fixed point $(0,0,0,0)$, as a function of $r_0$,  for $B=1$.
 The blue (red) points refer to the $x_1$ ($x_3$) string configuration. The dashed line shows the MSS bound in Eq. \eqref{eq:1}: $\dd \lambda_{MSS}=2 r_h \left(1 -\frac{B^2}{6 r_h^4}\right)$.}
\label{fig:lyapunovr0}
\end{center}
\end{figure}

\begin{figure}[b!]
    \begin{center}
    \includegraphics[width=0.5 \textwidth]{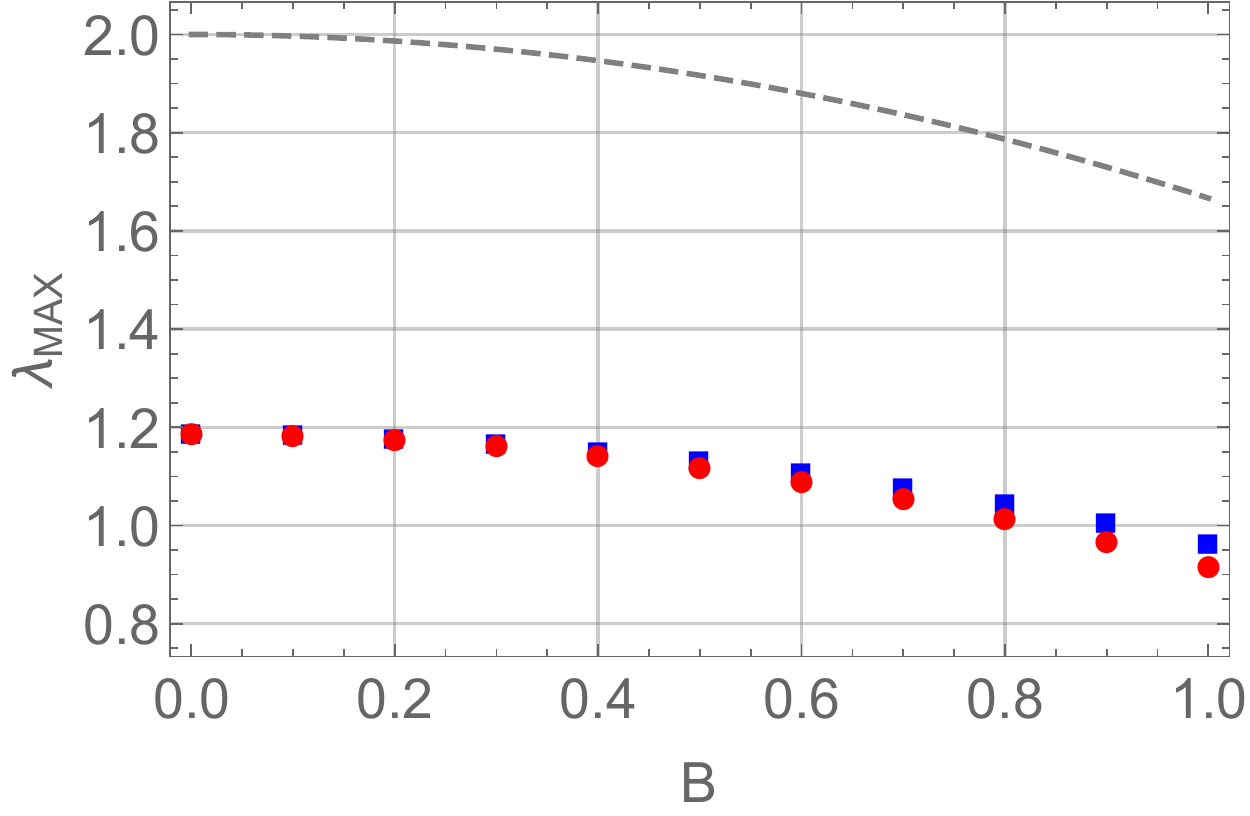}
    \caption{\small Largest Lyapunov exponent computed at the unstable fixed point $(0,0,0,0)$ as a function of $B$ for  $r_0=1.1$. The symbols are as in Fig.~\ref{fig:lyapunovr0}.}
    \label{fig:lyapunovB}
    \end{center}
    \end{figure}

\begin{figure}[h!]
\begin{center}
\includegraphics[width=0.5 \textwidth]{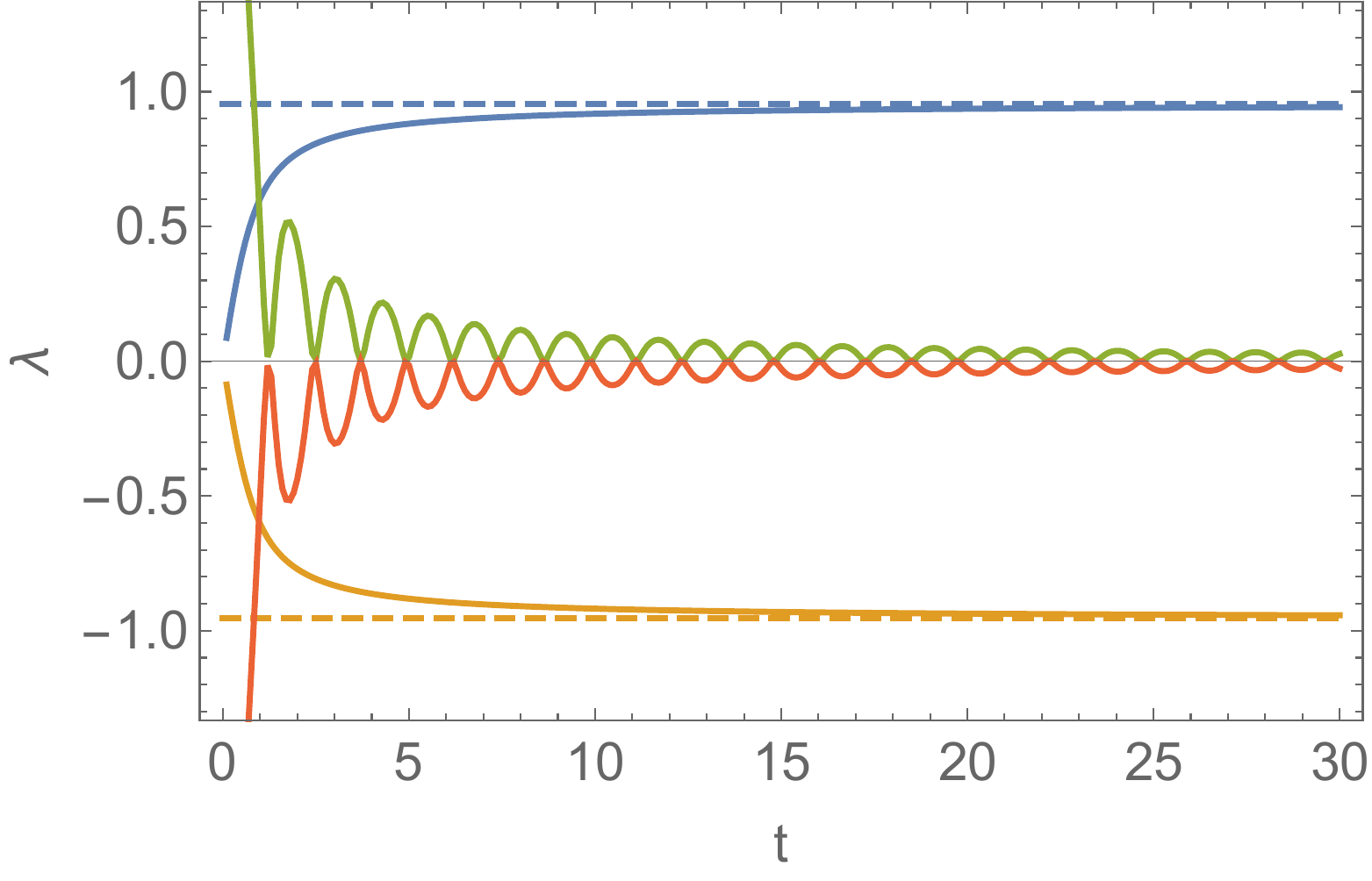}
\caption{\small Convergency plot of the  Lyapunov exponents for $r_0=1.1$ and $B=1$ computed at the unstable fixed point $(0,0,0,0)$. The horizontal lines correspond to the analytical values of the nonvanishing Lyapunov exponents $\sqrt{-\omega_0^2}$ (top  dashed blue line) and $-\sqrt{-\omega_0^2}$ (bottom  dashed orange line).}
\label{fig:globallyapunov}
\end{center}
\end{figure}

If an unstable unperturbed string is chosen as solution of Eqs.~\eqref{eq:eqxstatic}-\eqref{eq:eqrstatic}, for the  energy $E=0$ the unstable fixed point is at $\vec x=(0,0,0,0)$. 
At a fixed point the formula defining the Lyapunov exponents gets simpler \cite{sandri1996numerical}, and they can be computed analytically as the real part of the eigenvalues of the Jacobian matrix of $\vec F$.
At the point $(0,0,0,0)$ this Jacobian reads: 
\begin{equation}
J = \left(
\begin{array}{cccc}
0 & -\omega_0^2 & 0 & 0\\
1 & 0 & 0 & 0\\
0 & 0 & 0 & -\omega_1^2\\
0 & 0 & 1 & 0
\end{array}
\right)\,,
\end{equation}
and its eigenvalues  are $(-i \sqrt{\omega_0^2}, i \sqrt{\omega_0^2}, -i \sqrt{\omega_1^2}, i \sqrt{\omega_1^2})$. Hence, the Lyapunov exponents vanish for
 $\omega_i^2>0$, as for stable solutions.
Fig.~\ref{fig:lyapunovr0}  shows the largest Lyapunov exponents $\lambda_{MAX}=\sqrt{-\omega_0^2}$ at the fixed point $(0,0,0,0)$, for $B=1$ and varying  $r_0$. In Fig.~\ref{fig:lyapunovB} the exponents are plotted as a function of $B$  for $r_0=1.1$. 
The values of the Lyapunov exponent at this point are large, but they remain below the MSS bound $\lambda_{MSS}=2 r_h \left(1 -\frac{B^2}{6 r_h^4}\right)$ (dashed line in the plots) when the string tip gets close to the black hole horizon, $r_0\to r_h$. 
As a final check, in Fig.~\ref{fig:globallyapunov} it is shown that the Lyapunov exponents computed at the fixed point $(0,0,0,0)$ by the numerical procedure of Ref. \cite{sandri1996numerical} used in the previous section, are equal to the analytical ones $(\sqrt{-\omega_0^2}, -\sqrt{-\omega_0^2}, 0, 0)$.

\section{Conclusions}
Our investigation of a suspended string in a gravitational background with a black hole,  the holographic dual of the heavy quark-antiquark system in a thermal environment, confirms the MSS bounds \eqref{eq:1}  also in the case of  a uniform and constant magnetic field. The system becomes less chaotic increasing  $B$.  The anisotropy effect in two different orientations of the string is found.
This  conclusion is analogous to the one obtained  for different geometries, namely AdS-RN  \cite{Colangelo:2020tpr}, as well as studying the charged particle motion in such a kind of background \cite{Ageev:2018msv}.
Chaos has been  observed in the Poincar\'e plots, characterized by scattered points in the region close to the black-hole horizon, and quantitatively described  computing the  Lyapunov exponents, finding that the largest one verifies the MSS bound.
The stabilization effect of the magnetic field  is stronger for the string endpoints  lying on a line parallel to the field, keeping the black-hole horizon and the position of the tip of the string fixed. 
The largest Lyapunov exponents are below the MSS bound  also at the fixed unstable point of the potential describing the perturbed string   approaching  the horizon.

\vspace*{1.cm}
\noindent {\bf Acknowledgements.}
We thank F. De Fazio and S. Nicotri for discussions.
This study has been  carried out within the INFN project (Iniziativa Specifica) QFT-HEP.

\bibliographystyle{JHEP}
\bibliography{FNP1}

\end{document}